\documentclass[preprint,aps,floatfix]{revtex4} 

\usepackage{graphicx}
\usepackage{dcolumn}
\usepackage{bm}
\usepackage{longtable}
 \usepackage{amsmath}
 \usepackage{amssymb}
\usepackage{float}

\begin{document}


\title{Instrumentational complexity of music genres and why simplicity sells}

\author{Gamaliel Percino$^1$, Peter Klimek$^1$, Stefan Thurner$^{1,2,3}$}
\email{stefan.thurner@meduniwien.ac.at}
\affiliation{
$^1$Section for Science of Complex Systems; CeMSIIS; Medical University of Vienna; Spitalgasse 23; A-1090; Austria. \\
$^2$Santa Fe Institute; 1399 Hyde Park Road; Santa Fe; NM 87501; USA. \\
$^3$IIASA, Schlossplatz 1, A-2361 Laxenburg; Austria.
}%

\begin{abstract}
Listening habits are strongly influenced by two opposing aspects, the desire for variety and the demand for uniformity in music.
In this work we quantify these two notions in terms of musical instrumentation and production technologies that are typically involved in crafting popular music.
We assign a ``complexity value'' to each music style.
A style is complex if it shows the property of having both high variety and low uniformity in instrumentation.
We find a strong inverse relation between variety and uniformity of music styles that is remarkably stable over the last half century.
Individual styles, however, show dramatic changes in their ``complexity'' during that period.
Styles like ``new wave'' or ``disco'' quickly climbed towards higher complexity in the 70s and fell back to low complexity levels shortly afterwards, whereas styles like ``folk rock'' remained at constant high complexity levels.
We show that changes in the complexity of a style are related to its number of sales and to the number of artists contributing to that style.
As a style attracts a growing number of artists, its instrumentational variety usually increases.
At the same time the instrumentational uniformity of a style decreases, i.e. a unique stylistic and increasingly complex expression pattern emerges.
In contrast, album sales of a given style typically increase with decreasing complexity.
This can be interpreted as music becoming increasingly formulaic once commercial or mainstream success sets in.
\end{abstract}

\maketitle

\section{Introduction}

The composer Arnold Sch\"onberg held that joy or excitement in listening to music originates from the struggle between two opposing impulses, "the demand for repetition of pleasant stimuli, and the opposing desire for variety, for change, for a new stimulus." \cite{Schonberg}.
These two driving forces -- the demand for repetition or uniformity and the desire for variety -- influence not only how we perceive popular music, but also how it is produced.
This can be seen e.g. in one of last year's critically most acclaimed albums, Daft Punk's {\it Random Access Memories}.
At the beginning of the production process of the album the duo behind Daft Punk felt that the electronic music genre was in its 'comfort zone and not moving one inch' \cite{RollingStone}.
They attributed this 'identity crisis' to the fact that artists in this genre mostly miss the tools to create original sounds and rely too heavily on computers with the same libraries of sounds and preset banks \cite{Billboard}.
{\it Random Access Memories} was finally produced with the help of 27 other featured artists or exceptional session musicians, who were asked to play riffs and individual patterns to give the duo a vast library to select from \cite{RAM}.
The percussionist stated that he used 'every drum he owns' on the album; there is also a track composed of over 250 different elements.
The record was awarded the 'Album of the Year 2013' Grammy and received a Metacritic review of 'universal acclaim' for, e.g. 'breath[ing] life into the safe music that dominates today's charts' \cite{MetaCritic}.
However, the best-selling album of 2013 in the US was not from Daft Punk, but {\it The 20/20 Experience} by Justin Timberlake.
The producer of this album, Timothy Mosley, contributed 25 Billboard Top 40 singles between 2005-2010, more than any other producer \cite{HiphopDX}.
All these records featured  a unique production style consisting of 'vocal sounds imitating turntable scratching, quick keyboard arabesques, grunts as percussion' \cite{NYTimes}.
Asked about his target audience, Mosley said 'I know where my bread and butter is at. [...] I did this research. It's the women who watch Sex and the City' \cite{MTVNews}.
These two anecdotes illustrate how Sch\"onberg's two opposing forces, the demand for both uniformity and variety, influence the crafting of popular music.
The Daft Punk example suggests that innovation and increased variety is closely linked to the involved musicians' skills and thereby to novel production tools and technologies.
The example of Mosley shows how uniformity in stylistic expressions can satisfy listener demands and produce large sales numbers over an extended period of time.

The {\it complexity} of a style is determined by the set of specialized skills that are typically required of musicians to play that style.
Complexity of a style increases with (i) the number of skills required for the style and (ii) the degree of specialization of these skills.
A highly complex music style requires a diverse set of skills that are only available to a small number of other styles.
A style of low complexity requires only a small set of generic and ubiquitous skills, that can be found in a large number of other styles.
While it is hard to quantify skills of musicians and their capabilities directly, they can be related to the instrumentation typically used within a given music style.
If a music style requires a highly diverse set of skills, this will to some degree also be reflected in a higher number of different instruments and production technologies.
In general, demand for variety translates into a larger number of instruments used in the production process.
More diverse instrumentation indicates a higher number of skills of the involved musicians.
Desire for uniformity favors a limited variability in instrumentation in a production.
Music styles with high complexity therefore have large instrumental variety and at the same time low uniformity.
It follows that the desire for variety and uniformity are not only relevant for the perception of musical patterns.
The notions of variety and uniformity also apply to the involved capabilities of musicians and the instrumentations for their pieces.

Progress in a quantitative understanding of systemic trends in the music industry has been fueled by the recent availability and growth of online music databases, such as {\it www.discogs.com} or {\it www.allmusic.com}.
This data allows to uncover basic properties of collaboration and networks of artists \cite{Cano1, Cano3}, or to map song moods and listening habits of a large group of people \cite{Cano2, Lambiotte1}. 
Music data has been used to classify music genres \cite{Sordo,Buda2} through percolation techniques based on user annotations of music records \cite{Lambiotte2}, and to understand the emergence of pop stars \cite{Buda1}.
It has been shown that big cities tend to create more innovative styles \cite{Conrad}.
Recently an experiment was designed to determine the unpredictability of the success of a song without taking into account traditional marketing strategies such as broadcasting \cite{Watts}. 
It has been argued that audiences prefer music of intermediate complexity, the so-called ``optimal complexity hypothesis'' \cite{North}.
In \cite{Serra} popular music over the last fifty years was studied by analyzing pitch, timbre, and loudness of records.
Timbre, an attribute of the quality of sound, and {\it the} fingerprint of musical instruments, was found to experience growing homogenization over time. 

In this work we quantify the variety and uniformity of music styles in terms of instrumentation that is typically used for their production.
We employ a user-generated music taxonomy where albums are classified as belonging to one of fifteen different music genres which contain 374 different music styles as subcategories.
We show that styles belonging to the same music genre are characterized by similar instrumentation.
We use this fact to construct a similarity network of styles, whose branches are identified as music genres.
We characterize  the complexity of each music style by its variety and uniformity and show (i) that there is a remarkable relationship between varieties and uniformities of music styles, 
(ii) that the complexity of individual styles may exhibit dramatic changes across the past fifty years, and
(iii) that these changes in complexity are related to the typical sales numbers of the music style.

\section{Results}

\subsection{Music styles and genres are characterized by their use of instruments}
We introduce a time-dependent bipartite network connecting music styles to those instruments that are typically used in that style.
The dataset contains music albums and information on which artists are featured in the album, which instruments these artists play, the release date of the album, and the classifications of music genres and styles of the album.
For more information see the methods section and Tab. S1.

\begin{figure}[H]
  \centering
  \includegraphics[width=\textwidth]{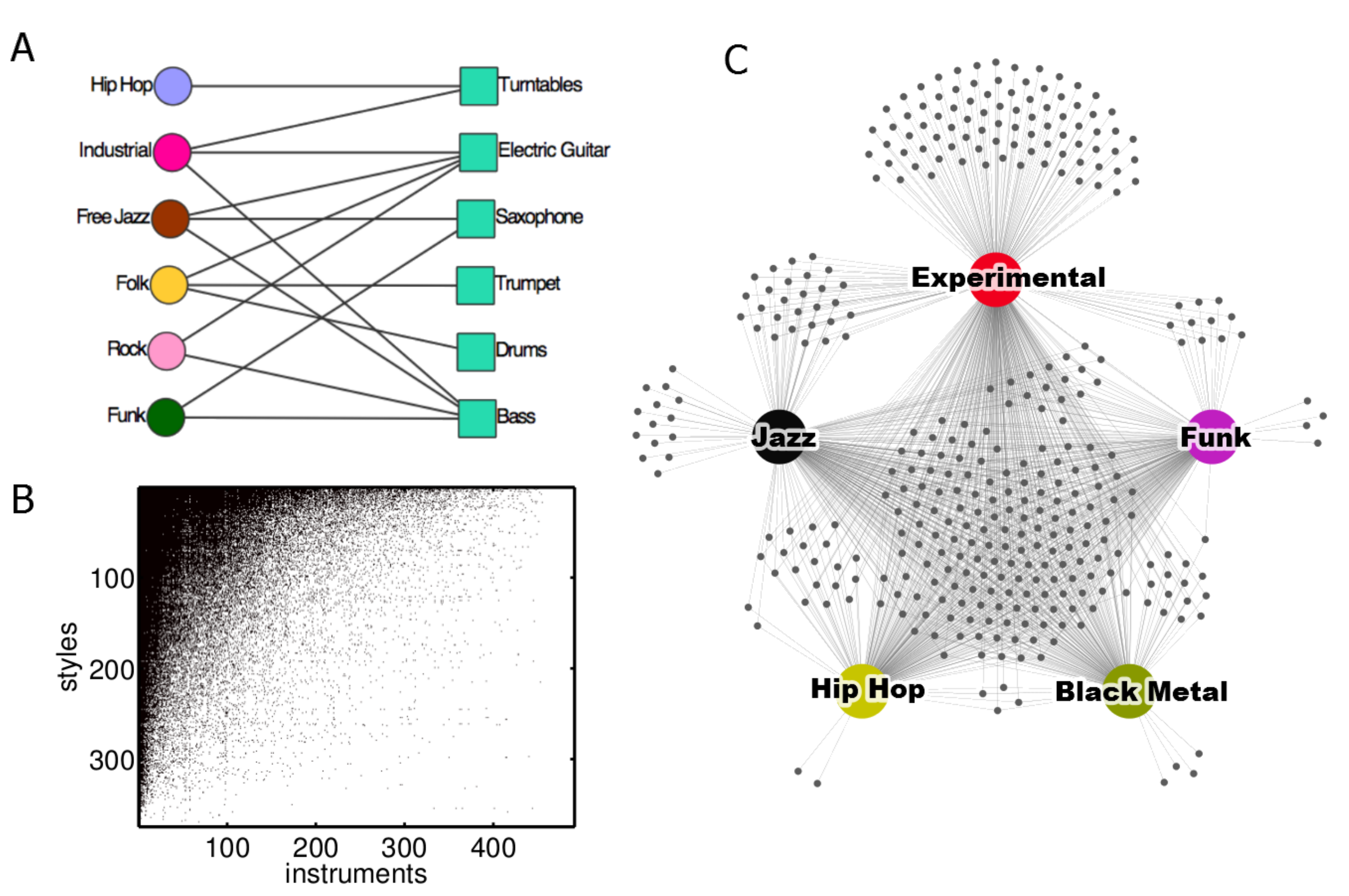}
  \caption
   {(A) Schematic representation of the data containing the relations between styles and instruments. (B) Visualization of the matrix describing the music production network, $M(t)$. A black (white) field for style $s$ and instrument $i$ indicates that $M_{si}(t) = 1(0)$. (C) Part of the bipartite network $M(t)$ that connects music styles with musical instruments for a given year $t$. Large nodes represent music styles, small ones instruments. It is apparent that some instruments occur in almost every style while others are used by a substantially smaller number of styles. For instance, there are only two instruments appearing exclusively in ``hip hop'' whereas dozens of instruments are only related to ``experimental''.}
   \label{fig:SIbip}
\end{figure}

We use the following notation.
If an album is released in a given year $t$, if it is classified as music style $s$, and contains the instrument $i$, this is captured in the time-dependent {\it music production network} $M(t)$, by setting the corresponding matrix element to one, $M_{si}(t) = 1$.
Figure \ref{fig:SIbip}A shows a schematic representation of the relations between several instruments and styles, and Fig. \ref{fig:SIbip}B shows the music production network $M(t)$ for the year 2004-2010. 
If instrument $i$ does not occur in any of the albums assigned to style $s$ released in time $t$, the matrix element is zero,  $M_{si}(t) = 0$. 
Let $N(s,t)$ be the number of albums of style $s$ released at time $t$. 
We only include styles with at least $h$ albums released within a given time window, $N(s,t)\geq h$. 
If not indicated otherwise, we choose $h=50$. 
The music production network $M_{si}(t)$ can be visualized as a dynamic bipartite network connecting music styles with instruments.
Figure \ref{fig:SIbip}C shows a snapshot of this bipartite network for five different music styles.

\begin{figure}[ht]
  \includegraphics[width=0.8\textwidth]{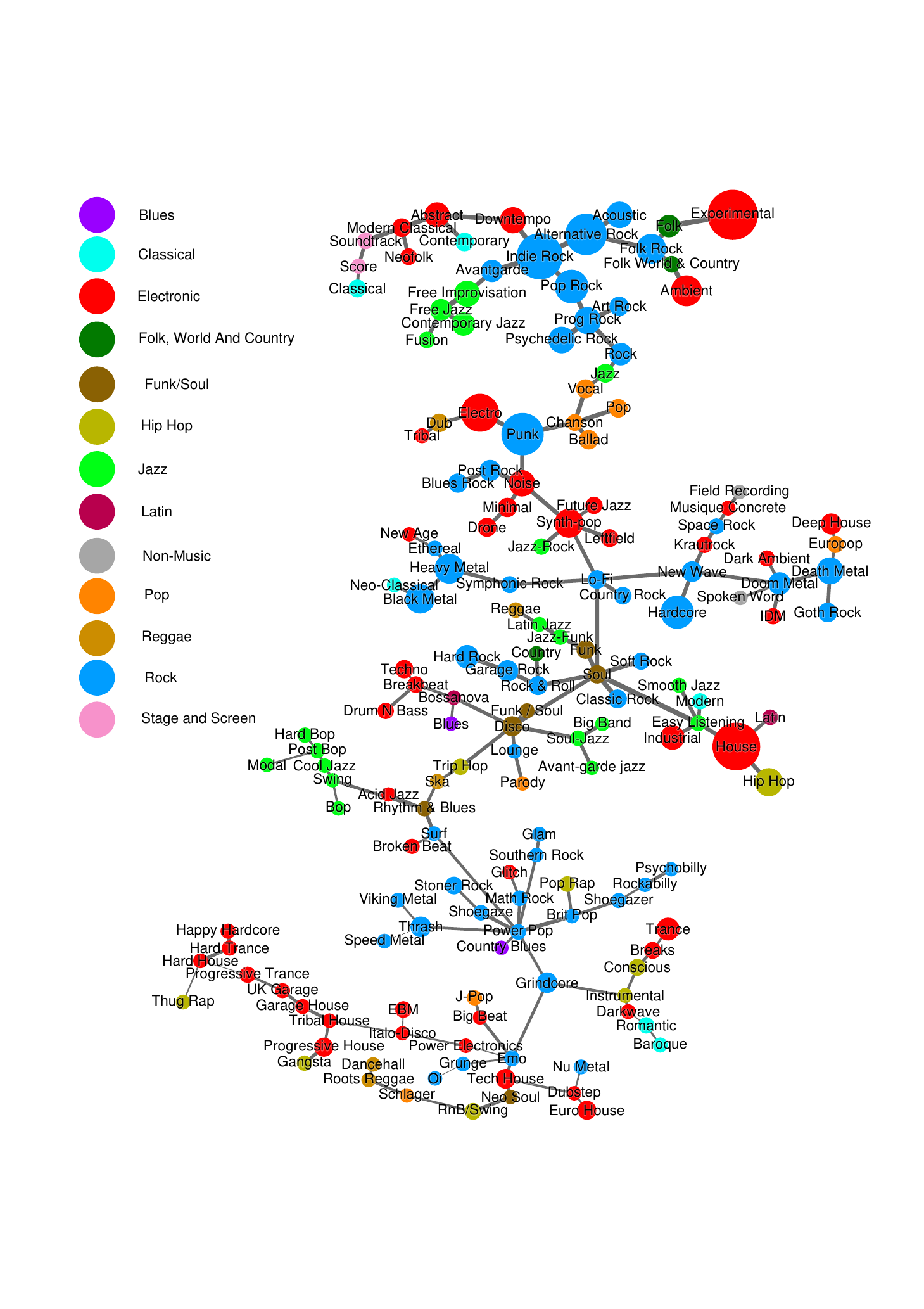}
 \caption
   {Maximum spanning tree for the style-similarity network $S$ for the years 2004-2010. Nodes represent styles, colors correspond to the genre to which the style belongs, the size is proportional to the number of albums released for each style, the link strength is proportional to $S_{s_1,s_2}(t_f)$. Several clusters are visible. They are identified as styles belonging to ``rock'', ``jazz'', or ``electronic music'' genres.}
  \label{fig:mst}
 \end{figure}

The similarity of two styles $s_1$ and $s_2$ can be computed by the overlap in instruments which characterize both styles at time $t$, as measured by the similarity network $S_{s_1,s_2}(t)$ that is defined in Methods.
Figure \ref{fig:mst} shows the maximum spanning tree (MST) of the style similarity network $S_{s_1,s_2}(t_f)$ computed for the last time period in the data, $t_{f}=$2004-2010.
The size of nodes (styles) is proportional to the number of albums $N_s(t_f)$ of style $s$ released in period $t_f$.
The data categorizes styles into genres, see Methods and supporting information Fig. S1.
Node colors indicate music genres, the strength of links is proportional to the value of $S_{s_1,s_2}(t_f)$. 
The MST shows several groups of closely related styles that belong to the same genres, such as ``rock'', ``electronic music'', or ``jazz''.
These clusters can be related to characteristic sets of instruments defining these genres.
``Jazz'' is mostly influenced by music instruments such as saxophone, trumpet and drums.
``Rock'' typically involves electric guitars, synthesizer, drums and keyboards, whereas ``electronic music'' is characterized by synthesizer, turntables, samplers, drum programming, and laptop computers. 

\subsection{Variety and uniformity define instrumentational complexity}
The {\it instrumentational variety} $V(s,t)$ of style $s$ at time $t$ is the number of instruments appearing in those albums that are assigned to $s$,
\begin{equation}\label{eq:eq1}
V(s,t)=  \sum_{i}M_{si}(t) \quad.
\end{equation}
$V(s,t)$ depends on how many different skills or capabilities of musicians (such as playing an instrument) are typically found within a music style.
{\it Instrumentational uniformity} $U(s,t)$ of style $s$ at time $t$ is the average number of styles that are related to an instrument that is linked to style $s$, or explicitly
\begin{equation}\label{eq:eq2}
U(s,t) = \frac{1}{\sum_{i}M_{si}(t)} \sum_{i} \left( M_{si}(t) \sum_{s'} M_{s'i}(t) \right) \quad  .
\end{equation}
To put it differently, the instrumentational uniformity of a given style $s$ is the average number of styles in which an instruments linked to $s$ is typically used.
Low (high) values of $U(s,t)$ indicate that the instruments characterizing style $s$ tend to be used in a small (large) number of other styles.

$V(s,t)$ and $U(s,t)$ measure different aspects of {\it instrumentational complexity}. 
These indicators are reminiscent of measures proposed  to quantify the complexity of economies of countries by the analysis of bipartite networks that connect countries to their exports of goods \cite{Hidalgo1}.
It was shown that changes in indicators resembling $V(s,t)$ and $U(s,t)$ are predictive for changes of national income.

As a measure for the number of sales of an album we use its Amazon ``SalesRank'', see Methods.
The average sales of a given music style $s$, $S(s)$, is given by the average SalesRank of albums assigned to style $s$,
\begin{equation} \label{eq:eq3}
S(s)= \langle SalesRank(r) \rangle_{r \in I(s)} \quad,
\end{equation}
where $I(s)$ is the index set of all albums $r$ that are assigned to style $s$. $S(s)$ is the average SalesRank of these albums.

\begin{figure}
  \centering
  \includegraphics[width=0.9\textwidth]{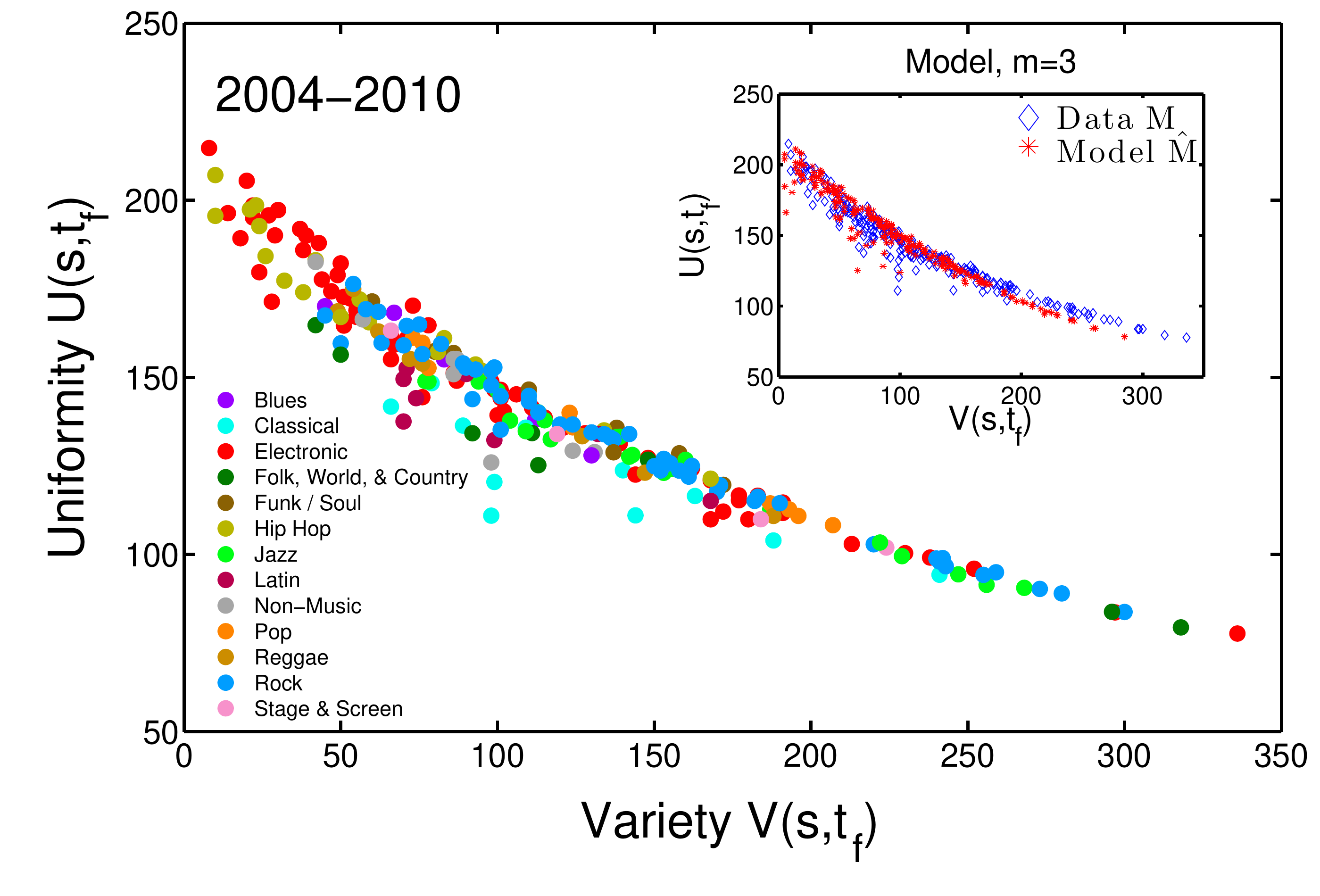}
  \caption
   {Instrumentational variety $V(s,t_f)$ and uniformity $U(s,t_f)$ for music styles within the time period $t_{f}$=2004-2010. Music styles collapse onto a line where variety and uniformity are inversely related. ``Experimental'' is the music style with the highest variety, styles with the lowest levels of variety and highest uniformity values belong to the ``electronic'' and ``hip hop'' genres. {\it Inset:} The values for instrumentational variety and uniformity are similar to results from the model.} 
 \label{fig:SIbip1}
\end{figure}

{\it Instrumentational complexity} of a music style can be expressed as the property of having high variety and low uniformity, i.e. the music is produced with a large number of different instruments which only appear in a small number of other styles.
Such production processes require musicians with a diverse and highly specialized set of skills.
As a complexity index $C(s,t)$ of a style $s$ at time $t$ we use
\begin{equation}
C(s,t) = \frac{V(s,t)}{U(s,t)} \quad.
\label{defC}
\end{equation}
Figure \ref{fig:SIbip1} shows each style (containing at least 50 albums) at time $t_f$ in the $V(s,t_f)$-$U(s,t_f)$ plane.
The styles follow a particular regularity: the higher the variety $V(s,t_f)$ of a music style, the lower its uniformity $U(s,t_f)$.
The style with the highest variety is ``experimental'', a style that categorizes music that goes beyond the frontiers of well established stylistic expressions. 
Most of the 20 styles with highest variety ($V(s,t_f)>230$) belong to the ``rock'' genre. 
Styles with low variety ($V(s,t_f)<75$) mostly belong to the ``electronic'' and ``hip hop'' genres. 
Interestingly, styles that deviate most from the curved line in Fig. \ref{fig:SIbip1} by having a comparably low uniformity correspond to styles such as ``Medieval'', ``Renaissance'', ``Baroque'', ``Religious'', and ``Celtic''.
These styles are played using unique instruments that require musicians with special training. 
In Fig. \ref{fig:SIbip1} the styles with high complexity can be found in the lower right quadrant of the plot, whereas styles with low complexity populate the upper left quadrant.

\subsection{Complexity-lifecycles of music styles}

\begin{figure}[!t]
  \centering
  \includegraphics[width=0.5\textwidth]{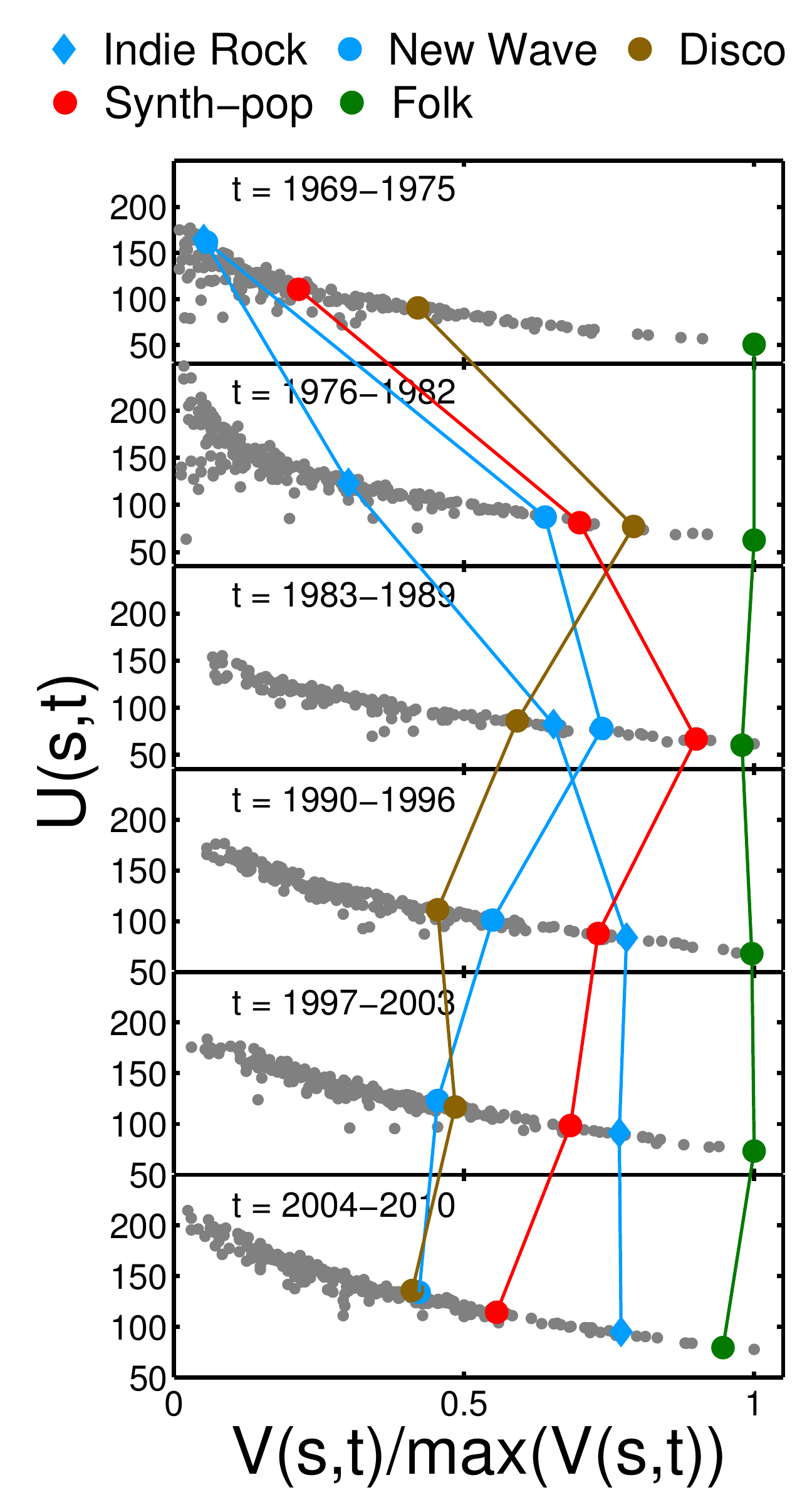}
  \caption
   {The arrangement of styles in the $V$-$U$ plane remains robust over more than fifty years of music history. However, the position of individual styles can change dramatically over time, as it is shown for ``indie rock'', ``new wave'',  ``disco'' and ``synth-pop''. Some styles, such as ``folk'', show almost no change in their position. }
   \label{fig:TimeRank}
\end{figure}

The relationship between instrumental variety and uniformity of styles is remarkably stable over time.
Variety $V(s,t)$ and uniformity $U(s,t)$ have been computed for six time-windows of seven years, starting with $t$=1969-1975.
For each time period $V(s,t)$ and $U(s,t)$ show a negative relation in Fig. \ref{fig:TimeRank}.
Values of $V(s,t)$ are normalized by $\frac{V(s,t)}{\max(V(s,t))}$ to make them comparable across time.
Although this relation is stable over time, the position of individual styles within the plane can change dramatically, as can be seen in the highlighted trajectories of several styles.
The evolution of music styles is also shown in the supporting Fig. S2 where the trajectory of $C(s,t)$ is shown for each style that ranks among the top 20 high complexity styles.
For example, the style ``new wave'' sharply increased in complexity rapidly and was popular from the mid-70's to the mid-80's, after which it decreased again.
Similar patterns of rise and fall in complexity are found for ``disco'' and ``synth-pop'' music.
``Indie rock'' gained complexity steadily from the 60s to the 80s and remained on high complexity levels ever since. 
Styles losing complexity over time include ``soul'', ``funk'', ``classic rock'', and ``jazz-funk''.
However, other styles such as ``folk'', ``folk rock'', ``folk world'', or ``country music'' remain practically at the same level of complexity.

\begin{figure}[!t]
  \centering
   \includegraphics[width=0.5\textwidth]{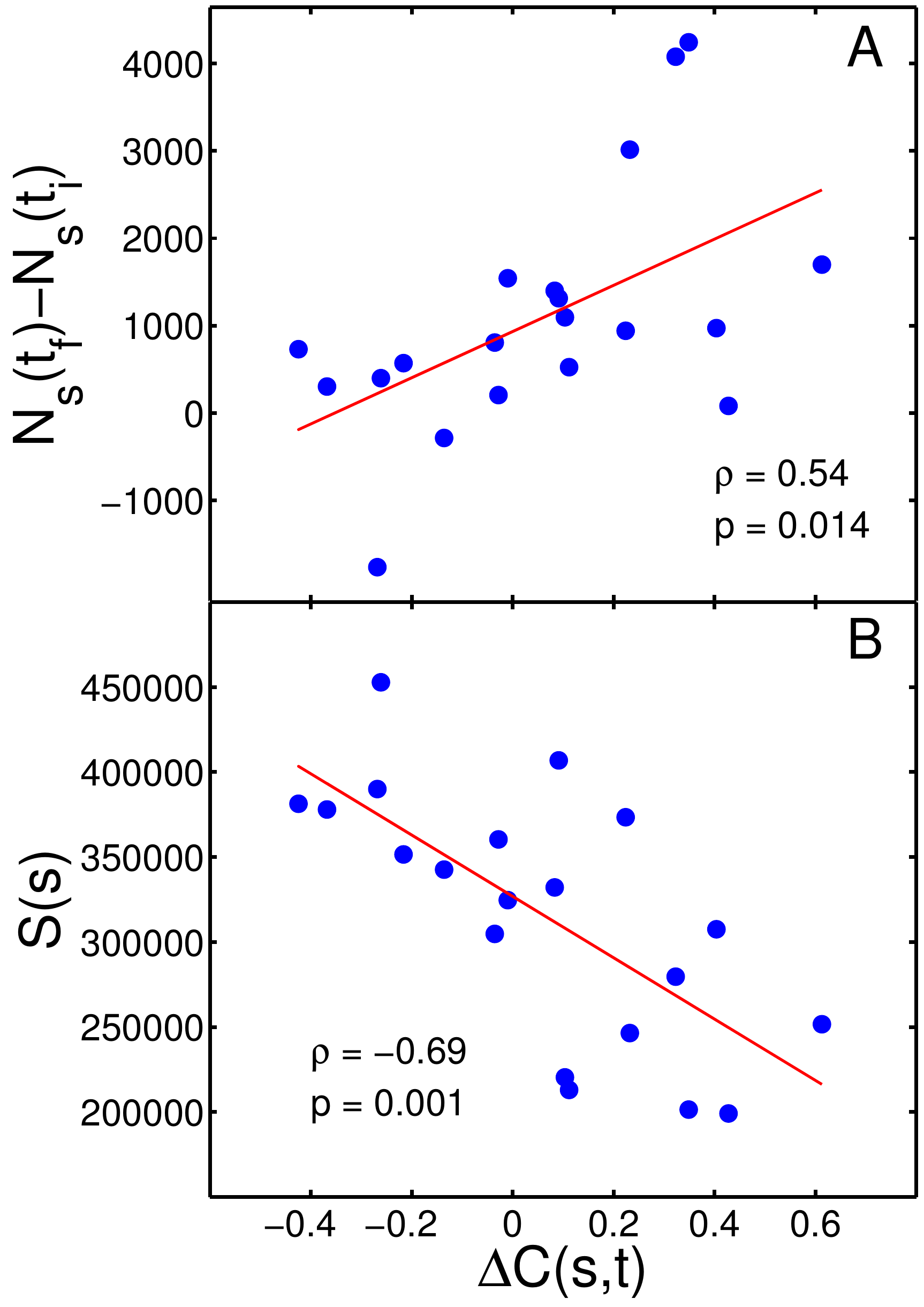}
  \caption
  {(A) Changes in number of albums $N_s$ versus changes in instrumentational complexity of music styles $\Delta C(s,t)$. We find a positive correlation with $\rho=0.54$ and p-value $p=0.0014$. (B) Sales $S(s)$ is negatively correlated with the change of complexity of music styles with correlation coefficient $\rho=-0.69$ and p-value $p=0.001$.}
   \label{fig:Sales}
\end{figure}

To understand the mechanisms leading to an increase or decrease in variety and uniformity we compute the change in the number of albums for each style between two seven-year windows, $t_{i}=$1997-2003 and $t_{f}=$2004-2010.
The change in number of albums is compared with changes in instrumentational complexity $\Delta C(s,t) = C(s,t_f) - C(s,t_i)$, see Fig. \ref{fig:Sales}A. 
We find that increasing complexity is correlated with an increasing number of albums within that time-span with a correlation coefficient $\rho = 0.54$ and p-value $p=0.014$.
This suggests that styles with increasing complexity attract an increasing number of artists that release albums within that style.

There exists a remarkable relation between changes in instrumentational complexity of a style and its average number of sales.
Figure \ref{fig:Sales}B shows that $\Delta C(s,t)$ is negatively correlated with the average number of sales $S(s)$ as defined in Eq. (\ref{eq:eq3}), with a correlation coefficient, $\rho=-0.69$ and a p-value, $p = 0.001$. 
Naively, one could assume that styles with increasing sales numbers show increasing numbers of albums, since they offer more prospect for generating economic revenue.
However, the opposite is true, $S(s)$ and the change in albums $N(s,t_f) - N(s,t_i)$ are negatively correlated with correlation coefficient $\rho= -0.46$ and p-value $p = 0.04$.
Supplementary Fig. S3 shows a version of Fig. \ref{fig:Sales} where each data point is labeled by its style.
Note that here we take into account only  styles $s$ that have at least 1,500 albums in periods $t_i$ and $t_f$ since only for those the average SalesRank can be estimated reliably.
For styles with more than 1,500 albums there are also no significant correlations between the number of albums per style $N(s,t)$ and the indicators $V(s,t)$ (correlation coefficient  $\rho=0.29$ and p-value $p=0.22$), $U(s,t)$ ($\rho=-0.29$, $p=0.22$), $C(s,t)$ ($\rho=0.26$, $p=0.28$), and $S(s)$ ($\rho=-0.17$, $p=0.5$).
It can therefore be ruled out that the results shown in Figs. \ref{fig:Sales}A and B are driven by changes in the number of albums for each style.
 
To summarize, the movement towards higher complexity of a style is correlated with a larger number of artists attracted to that style.
Musicians with highly specialized skills lead to increased variety and decreased uniformity.
A decrease in complexity is correlated with an increased number of album sales.
This can be interpreted in the following way.
As musicians experiment within a music style, thereby increasing its variety and decreasing its uniformity, they eventually find a 'formula for commercial success', i.e. a recipe that attracts large audiences.
As this formula is repeatedly used in other albums, variety decreases and uniformity of released albums increases, giving an overall decrease in music complexity of the style.

\subsection{A simple model}

High instrumentational complexity typically requires musicians with a diverse and highly specialized set of skills.
We now show that the results for instrumentational variety and uniformity can be understood with a simple model that explicitly takes into account the capabilities of artists.
Therefore we introduce two bipartite networks that can be extracted from the data: the style-artist network $P(t)$ and the artist-instrument network $Q(t)$. 
Entries in $P(t)$ and $Q(t)$ are zero by default.
If a given artist $a$ is listed in the credits of an album released at time $t$ and assigned to style $s$, we set $P_{sa}(t)=1$. 
If the artist $a$ plays instrument $i$ on an album released at $t$, we set $Q_{ai}(t)=1$.
In the model we assume that instrument $i$ is associated to style $s$, if there are at least $m$ artists which are both related to instrument $i$, and to style $s$.
The model music production network $\widehat M(m,t)$ is given by
\begin{equation}\label{eq:eq5}
 \widehat M(m,t) =
  \begin{cases}
    1 & \text{if } \sum_{a}P_{sa}(t) Q_{ai}(t) \geq m \\
   0       &  \text{otherwise} \quad.
  \end{cases}
\end{equation}
From $\widehat M(m,t)$ we compute the model variety $\widehat V(s,t)$ and model uniformity $\widehat U(s,t)$.
The optimal choice of the threshold $m$ is found by maximizing the goodness-of-fit between data and model.
To this end we use $n$ data bins $x_i$ for instrumentational variety with intervals of size one, $x_i-x_{i-1} = 1$.
We define the uniformity $u_i$ ($\widehat u_i$) for the data (model) as the average uniformity of all styles $s$ with variety $V(s,t)$ ($\widehat V(s,t)$) from the interval $\left[x_i,x_i+1 \right)$.
The average squared residuals $R(m) = \tfrac{1}{n} \sum_{i=1}^n (u_i - \widehat u_i)^2$ are then calculated for $m=1,\dots, 10$.
For $m=3$ the value of $R(m)$ assumes its minimum, revealing that the model explains the data best, if one assumes that an instrument $i$ can be associated with a style $s$, given that there are at least $m=3$ artists that are both related to style $s$, and instrument $i$.
Supplementary Fig. S4 shows a comparison of data and model for various choices of $m$ and $h$.
The results for model variety and uniformity for $m=3$ are shown in the inset of Fig. \ref{fig:SIbip1}.
For the bulk of styles, data and model are practically indistinguishable.

Note that the results shown in Figs. \ref{fig:SIbip1} and \ref{fig:Sales} can {\it not} be explained by trivial features of the data such as numbers of instruments or artists per style alone.
This is shown by introducing randomized versions of $M$, and $\widehat M$.
A randomization of $M$ is obtained by replacing each row in $M$ by a random permutation of its elements, we call it $M^{rand}$.
The varieties of each style are the same for $M$ and $M^{rand}$, but uniformities will change.
Results for the relationship between variety and uniformity when computed from $M^{rand}$ are shown in Fig. S5 for two different choices of the threshold $h$ (50 and 1,500, respectively).
Figure S5 shows that there is no correlation between variety and uniformity for either choice of $h$, and that the data can not be reproduced.
Note that for $M^{rand}$ the styles have similar uniformity values, independent of their variety.
The non-trivial relation between $V(s,t_f)$ and $U(s,t_f)$ in Fig. \ref{fig:SIbip1} is therefore driven by the differing uniformities of music styles and can {\it not} be explained by variety alone.

A randomized version of the model music production network $\widehat M$, $\widehat M^{rand}$, is obtained by replacing both the style-artist network $P(t)$ and the artist-instrument network $Q(t)$ by randomizations.
In these randomizations $P(t)$ ($Q(t)$) is replaced by a random matrix that has the same size and number of zeros and ones as $P(t)$ ($Q(t)$), and where each entry is nonzero with equal probability.
That is, each artist is assigned a randomly chosen set of instruments and styles while the total number of associations is fixed.
Figure S6 shows that the high overlap between data and model disappears under this randomization.
There is also no significant correlation between sales $S(s)$ and the change in complexity for the randomized music production network $\widehat M^{rand}$.
The relationships between variety, uniformity, and sales numbers for the various music styles can only be explained by taking the skills of musicians into account, i.e. who is able to play which instrument under which stylistic requirements.

\section{Discussion}

We quantified variety and uniformity of music styles over time in terms of the instruments that are typically involved in crafting popular music.
This allowed us to construct a style-similarity network where styles are linked if they are associated with similar sets of instruments.
Clusters of styles in this network correspond to music genres, such as ``rock'' or ``electronic music''.
Instrumentational complexity of a music style is the property of having both, high variety, and low uniformity.
We found a negative correlation between variety and uniformity of music styles that was remarkably stable over the last fifty years.
While the overall distribution of complexity over music styles is robust, the complexity of {\it individual} styles showed dramatic changes during that period.
Some styles like ``new wave'' or ``disco'' quickly climbed towards higher complexity and shortly afterwards fell back, other styles like ``folk rock'' stayed highly complex over the entire time period.
We finally showed that these changes in the complexity of a style correlate with its sales numbers and with how many artists the style attracts.
As a style increases its number of albums, i.e. attracts a growing number of artists, its variety also increases.
At the same time the style's uniformity becomes smaller, i.e. a unique stylistic and complex expression pattern emerges.
Album sales numbers of a style, however, typically increase with {\it decreasing} complexity.
This can be interpreted as music becoming increasingly formulaic in terms of instrumentation under increasing sales numbers due to a tendency to popularize music styles with low variety and musicians with similar skills.
Only a small number of styles in popular music manages to sustain a high level of complexity over an extended period of time.

\section{Data and Methods}

\subsection{Data}
The {\it Discogs} database is one of the largest online user-built music database specialized on music albums or discographies.
Users can upload information about music albums. 
A group of moderators assures correctness of the information.
Discogs is an open source database and publicly accessible via API or XML 
dump file released every month.
We use the dump file of November 2011 containing more than 500,000 artist and more than 500,000 albums assigned to 374 styles.
The data spans more than fifty years of music history, from 1955-2011.
Discogs uses a music taxonomy based on two levels, music genres and styles.
There are fifteen different genres, such as ``rock'', ``blues'', or ``Latin''. 
On the second level genres are divided into styles, for instance ``rock'' has 57 styles including ``art rock'', ``classic rock``, ``grunge'', etc.
``Latin'' contains $44$ different music styles such as ``cumbia'', ``cubano'', ``danzon'', etc.
Supporting Fig. S1 shows the histogram of the distribution of music styles per genres.
For each music album we extract information on the instruments played by artists, the release date of the record, and the music genres and styles assigned to the album.
The data is grouped into time windows of seven years, e.g. the last time-step contains data on albums released between 2004-2010, and so on.
Supplementary Table S1 provides some descriptive statistics of the dataset.

To measure the average sales numbers of music styles we use a dataset \cite{alda1, alda2} that contains information on the Amazon SalesRank of music albums.
It was collected in summer 2006.
The Amazon SalesRank can be thought of as a ranking of all records by the time-span since an item last sold \cite{Amazon}.
Albums in the Discogs dataset are assigned their Amazon SalesRank by matching album titles between the two datasets. 
As the Amazon SalesRank dataset only contains information on album titles, it was matched to entries in the Discogs dataset by choosing only albums whose title appears only once in both datasets.

\subsection{Style similarity network}
The {\it style similarity network} $S$ quantifies how similar two music styles are in terms of their instrumentation.
A weighted link in matrix $S$ that connects two musical styles, $s_1$ and $s_2$, is defined as the number of instruments they have in common, divided by the maximum value of their respective varieties $V(s_{1/2},t)$.
At a given time $t$, the entries in $S$ are given by
\begin{equation}\label{eq:eq4}
S_{s_1,s_2}(t) = \frac{\sum_{i}M_{s_1i}(t) M_{s_2i}(t)}{\max[ V(s_1,t), V(s_2,t) ]} \quad.
\end{equation}
To visualize the network of music styles we compute the maximum spanning tree (MST) for $S$.
We then follow the visualization strategy presented in \cite{Hidalgo2}.

\section{Acknowledgement}
P.K. was supported by EU FP7 project MULTIPLEX, No. 317532, and G.P. by the National Council for Science and Technology of Mexico with the scholarship number 202117.

\pagebreak
\begin{center}
{\huge Supporting Information}
\end{center}
\setcounter{section}{0}
\setcounter{figure}{0} 

\makeatletter
\renewcommand{\thetable}{S\arabic{table}}%
\renewcommand{\fnum@figure}{\figurename~S\thefigure}
\makeatother

This Document is supplementary information of the manuscript "Instrumentational complexity of music genres and why simplicity sells". It contains information about the data set analyzed in the main manuscript and supplementary figures of the data set and the models generated.

\section{Data}

Discogs.com is a comprehensive, user-built music database with the aim to provide cross-referenced discographies of all labels and artists.
As of April 2014, more than 189,000 people have contributed to this collection.
We work with an XML dump of the database from November 2011.
The number of data entries is shown in Tab. \ref{tab:Discogs1}.
The dataset includes more than half a million artists and albums spanning the years 1955-2011, as well as almost 500 instruments.

\begin{table}[h]
\centering 
\caption{Overview of data extracted from $Discogs.com$} 
\begin{tabular}{c r} 
\hline\hline
Data type  & \multicolumn{1}{c}{Number of entries} \\ [0.5ex] 
\hline 
Artists & 580060  \\ 
Albums & 536422 \\
Instruments & 491\\
Styles & 374 \\
Genres & 15 \\
Years & 1955-2011 \\[1ex] 
\hline 
\end{tabular}
\label{tab:Discogs1}
\end{table}

\begin{figure}[tbp]
  \centering
  \includegraphics[width=0.45\textwidth]{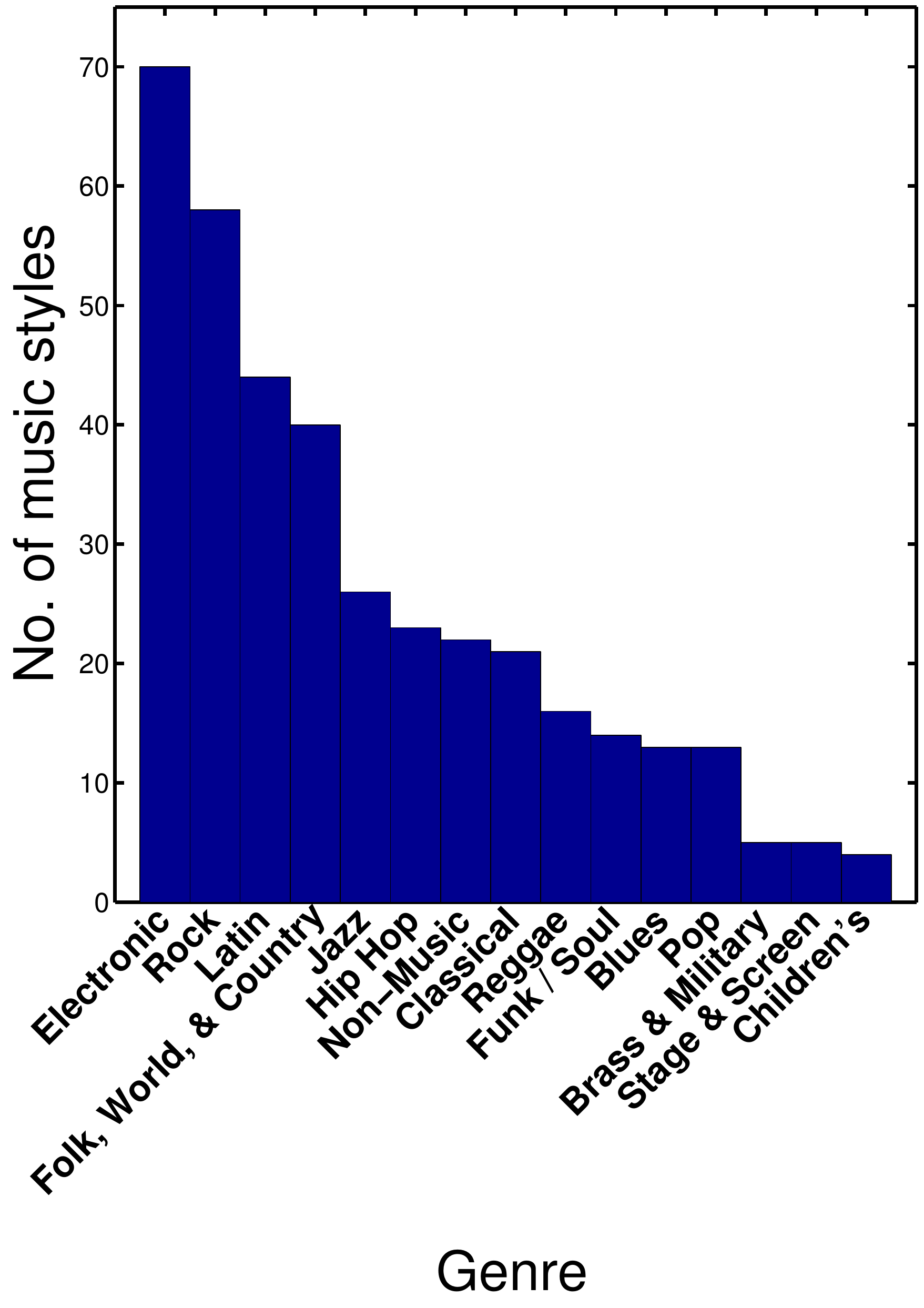}
  \caption
   {Rank-frequency plot of the number of styles per genre. ``Electronic'' and ``rock'' genres contain the largest numbers of styles, ``children's music'' the least number of styles. }
   \label{fig:HistGS}
\end{figure}

Discogs uses a music taxonomy for albums based on two levels.
On the first, highest level in the taxonomy there are 15 different music genres, for instance ``rock'', ``blues'', or ``electronic''.
On the second level the genres are broken down into 374 different styles, for example ``drum and bass'' is a style of the genre ``electronic''.
Fig. S\ref{fig:HistGS} shows a rank-frequency plot of the genres and the number of styles they contain.
The genres ``electronic'' and ``rock'' have the largest number of styles (more than fifty), whereas ``brass \& military'', ``stage \& screen'', and ``children's music'' have the smallest number of styles.
The genre and style information for albums is also entered by users, who may choose from a pre-specified list of styles.
This list of styles is generated by discogs users as the outcome of a collaborative and moderated process.

\section{Complexity life-cycles of music styles}

\begin{figure}[tbp]
  \centering
  \includegraphics[width=0.60\textwidth]{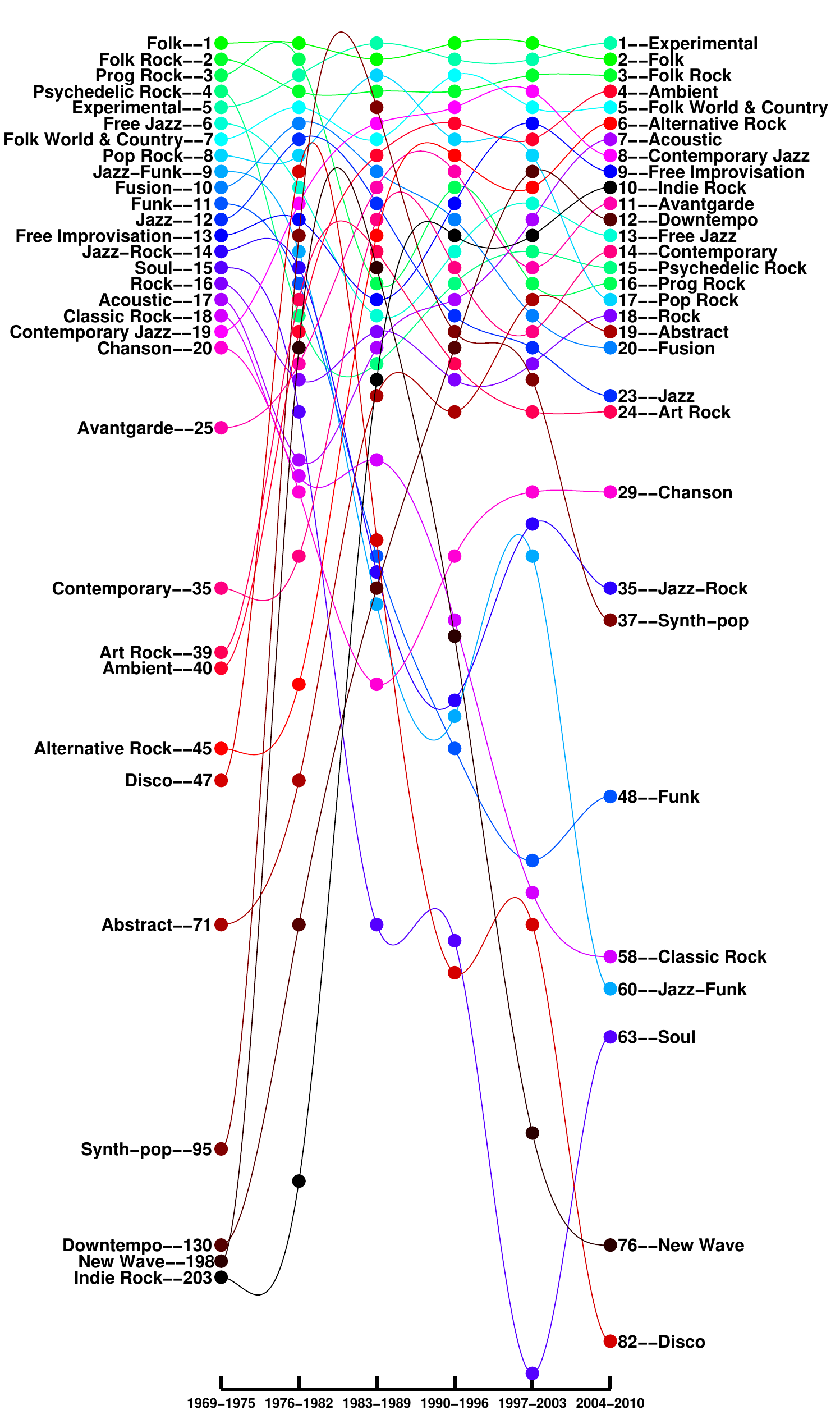}
  \caption
   {Styles are ranked according to their complexity in each of the studied time windows, and the changes in complexity are shown as trajectories for each style that ranked at least once among the top 20 in terms of complexity. ``Experimental music'', ``folk'' and ``country'' are nearly stationary, while ``indie rock'', ``new wave'', or ``disco'' changed their complexity-ranks dramatically.}
   \label{fig:Rank}
\end{figure}

Figure S\ref{fig:Rank} shows the trajectories for instrumentational complexity $C(s,t)$ for each style which was ranked at least once among the twenty styles with highest complexity.
``Experimental'', ``folk'', and ``folk rock'' rank among the top-5-variety-styles in each time window.
``New wave'' and ``indie rock'' start at a variety rank around 200 in 1969-1975 and show a stark increase in complexity over the next time-windows.
In 1983-1989 ``new wave'' reaches rank 5, ``indie rock'' is ranked \#15 in this time window.
However, afterwards their trajectories diverge drastically.
``New wave'' goes quickly down in complexity until it reaches a rank of 76 in 2004-2010, whereas ``indie rock'' continues to climb up to rank 10 in the last time window.
The styles ``disco'' and to a lesser extent ``synth-pop'' show the same pattern of complexity changes as ``new wave'', i.e. a rapid increase followed by an equally rapid decrease.
``Alternative rock'' and ``downtempo'' show complexity changes similar  to ``indie rock'', namely continual increases.
Other styles show a decline in complexity over time.
For example ``soul'', ``classic rock'' and ``funk'' have complexity ranks in the range 10-20 in 1969-1975, while non of them is ranked in the top 50 in 2004-2010.

\begin{figure}[tbp]
  \centering
  \includegraphics[width=0.5\textwidth]{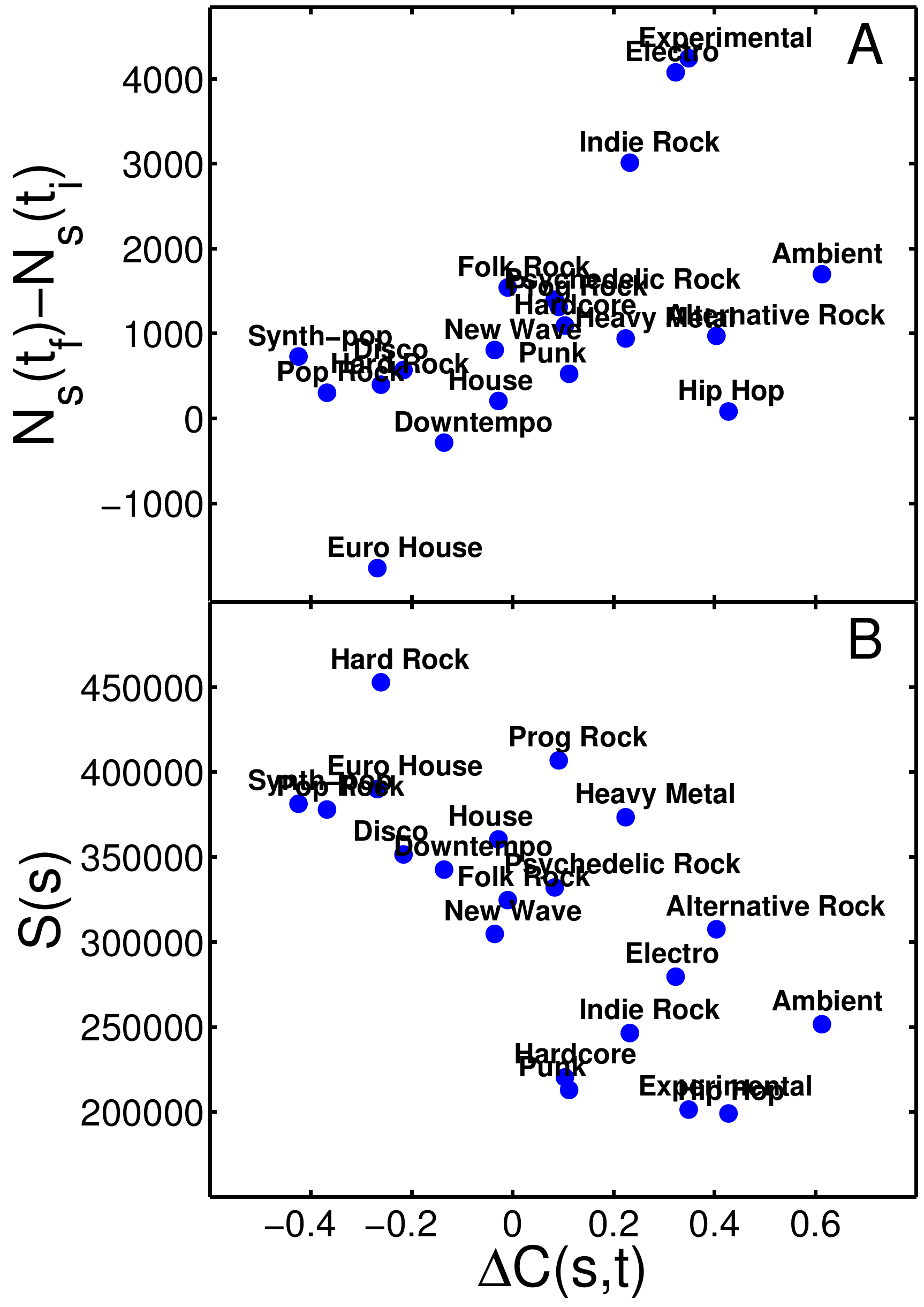}
  \caption
   {Same as Fig. 5 in the main text with labels for the music styles.}
   \label{fig:SalesRanStyles}
\end{figure}

Figure S\ref{fig:SalesRanStyles} shows a version of Fig. 5 from the main text with labels of the music styles for data points.
It becomes apparent that the styles ``euro house'', ``synth-pop'', ``disco'', ``pop rock'', and ``hard rock'' exhibit decreasing complexity, increased average sales numbers $S(s)$, and decreased numbers of albums.
``Experimental'', ``ambient'', ``alternative rock'', and ``hip hop'' show the largest increases in complexity over time, while their averages sales decrease and the number of related albums increases.

\begin{figure}[ht]
 \centering
  \includegraphics[width=0.7\textwidth]{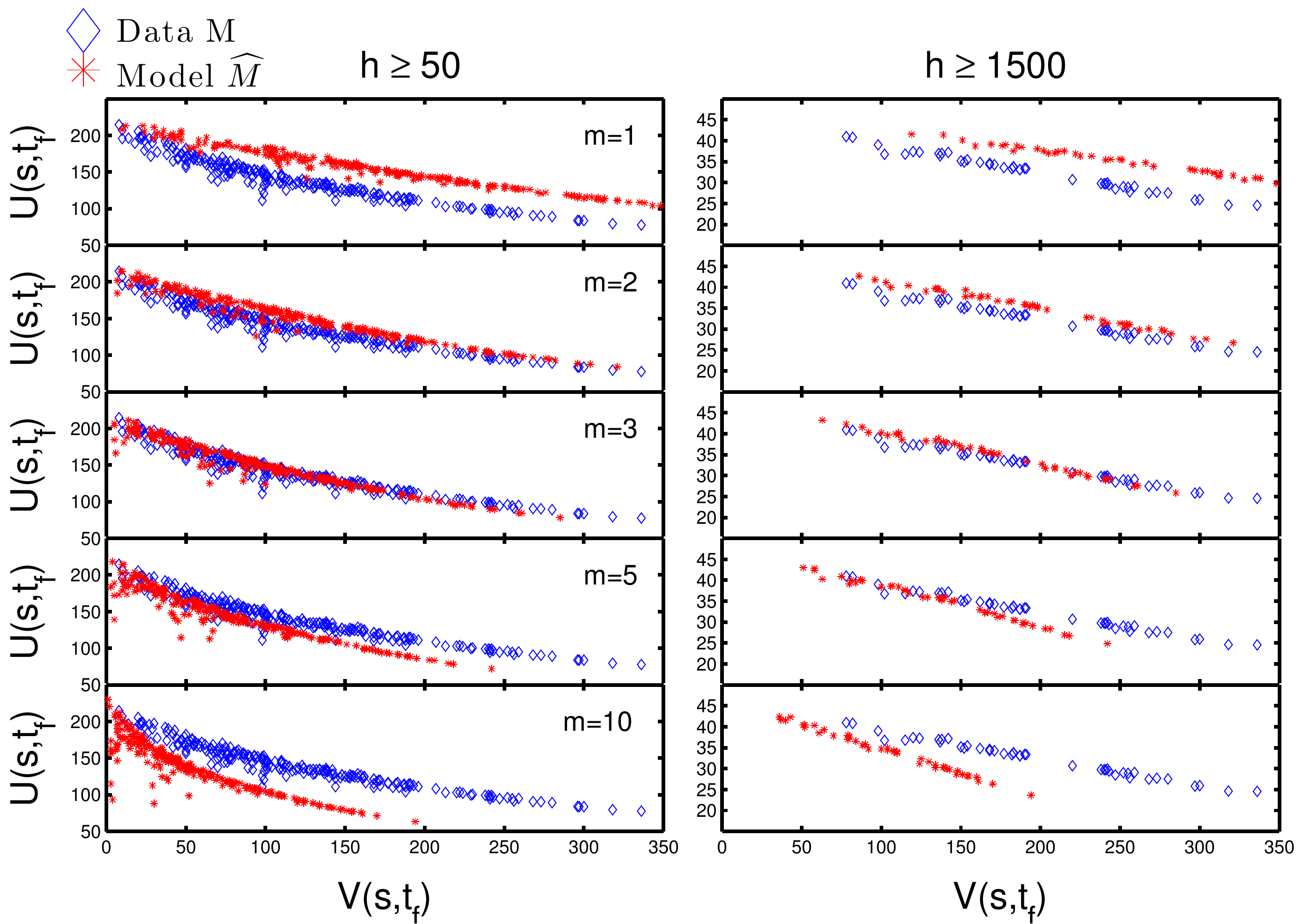}
  \includegraphics[width=0.7\textwidth]{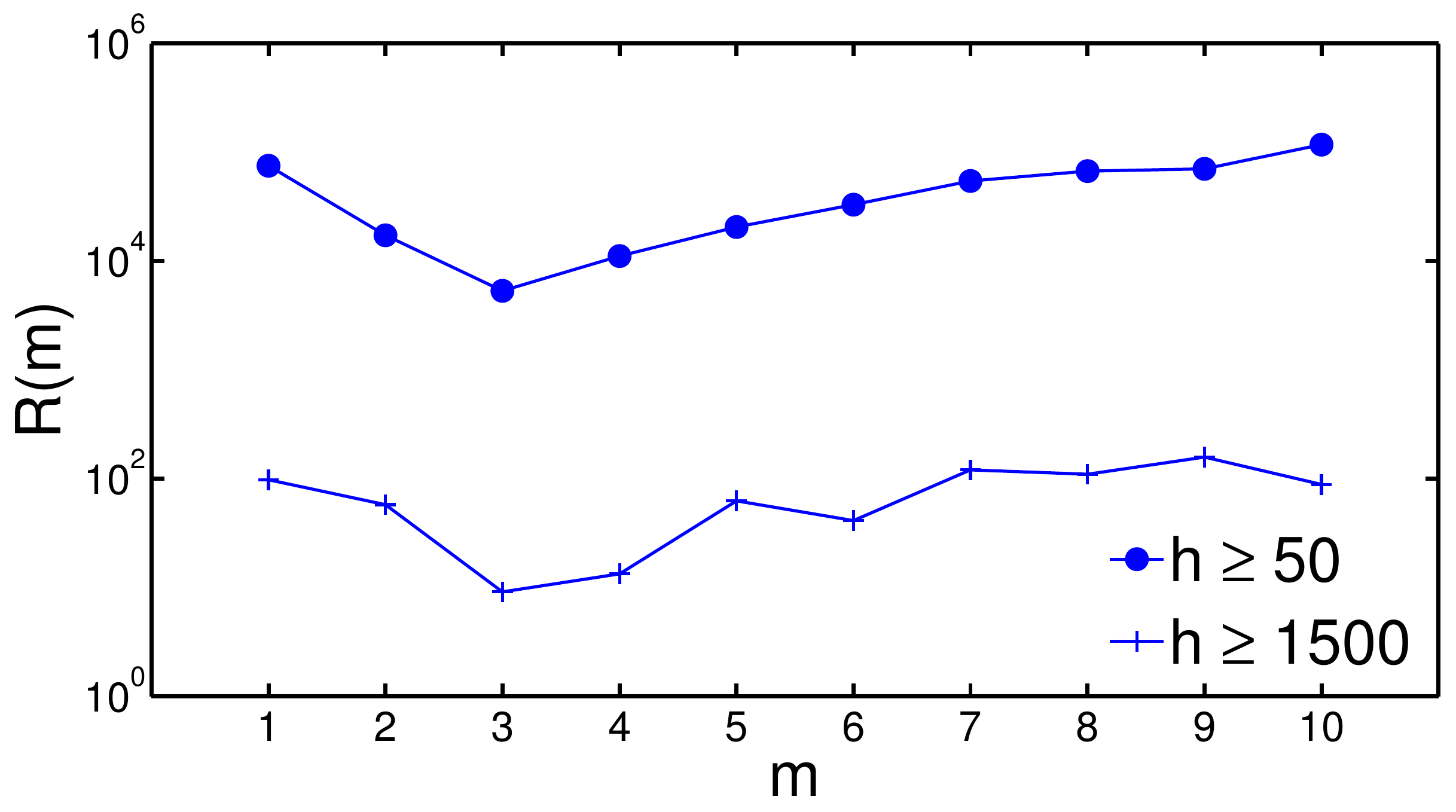}
  \caption
  {Variety and uniformity values for data and model for threshold values of $h=50$ (left) and $h=1500$ (right). Each row corresponds to a different value of $m=1,2,3,5,10$. The average squared residuals $R(m)$ are calculate for $m=1,...,10$ showing that for $m=3$ the model describes the data independently from the threshold $h$, meaning it is enough m=3 artist to associate and instrument $i$ with a style $s$. }
   \label{fig:Model2}
\end{figure}

Results for the distribution of styles in the instrumentational variety-uniformity plane are compared for data and model of Eq. (5) in Fig. S\ref{fig:Model2}.
In the left column of Fig. S\ref{fig:Model2} results for a threshold value of $h=50$ are shown, the right column shows results for $h=1500$.
Each row corresponds to a different value of $m=1,2,3,5,10$.
It is apparent that for both thresholds $h$ the highest overlap between data and model is found for $m=3$.
A model where it takes at least three musicians playing a given instruments and releasing albums in a given style, in order to constitute a style-instrument-relation, describes the data best.

\section{Randomization results}

\begin{figure}[h]
  \centering
  \includegraphics[width=0.45\textwidth]{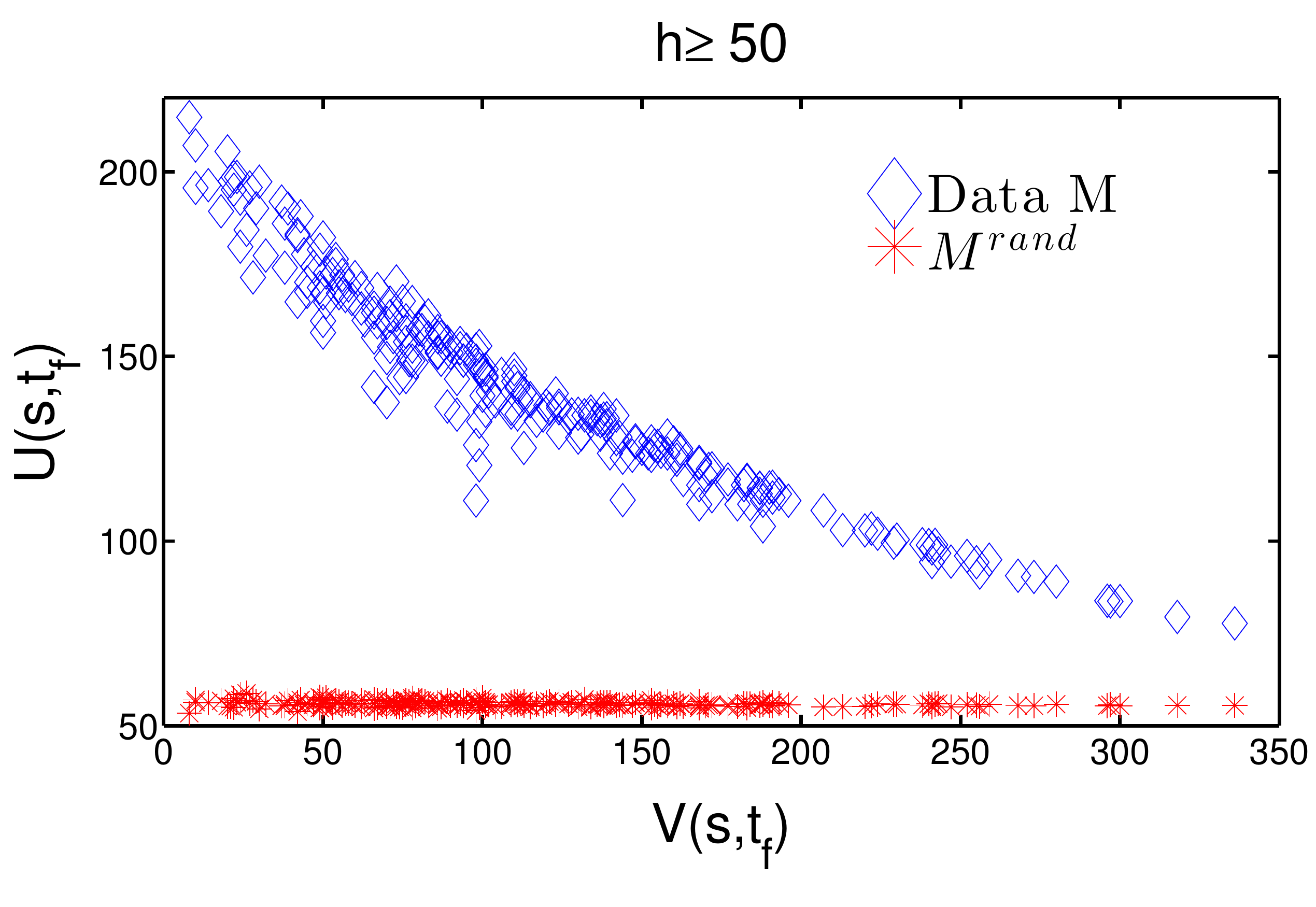}
  \includegraphics[width=0.45\textwidth]{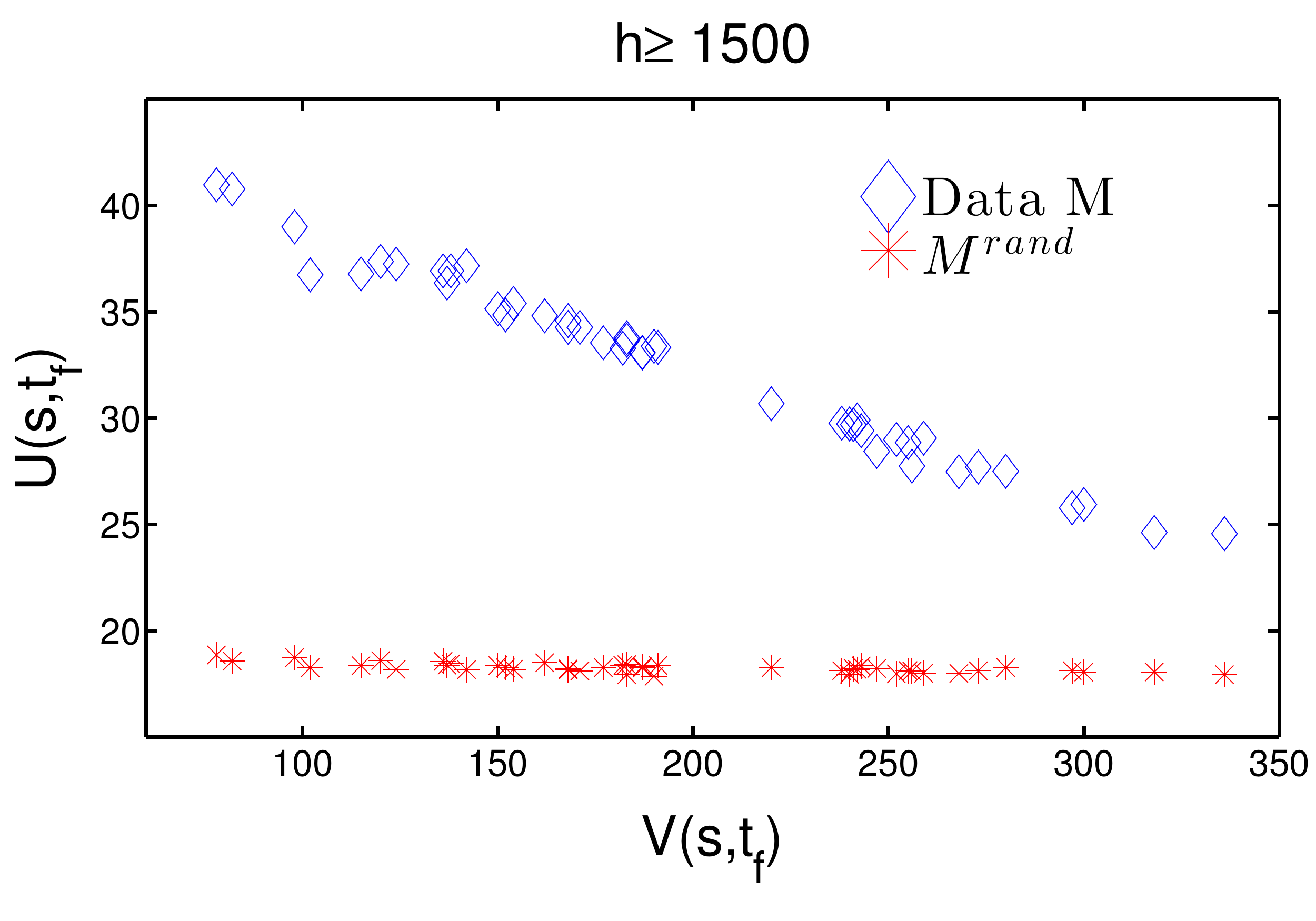}
  \caption
  {The distribution of styles in the $V-U$-plane computed from the data $M$ and its randomization $M^{rand}$ shows that the negative correlation between $V$ and $U$ is destroyed by randomizing the instruments associated to each style.  The relation between $V$ and $U$ is therefore the result of a nontrivial structure captured by the uniformities of styles.}
 \label{fig:Model1}
\end{figure}

The distribution of variety and uniformity values of music styles for the music production network $M$ obtained from the data is compared to the distribution of styles from the randomized production network $M^{rand}$ in Fig. S\ref{fig:Model1} for two threshold values, $h=50$ and $h=1500$.
The randomization destroys the negative correlation between variety and uniformity.
The styles have similar levels of uniformity, independent from their variety values, and the results for $M^{rand}$ resemble the results obtained from $M$ only very poorly.

\begin{figure}[t]
  \includegraphics[width=0.45\textwidth]{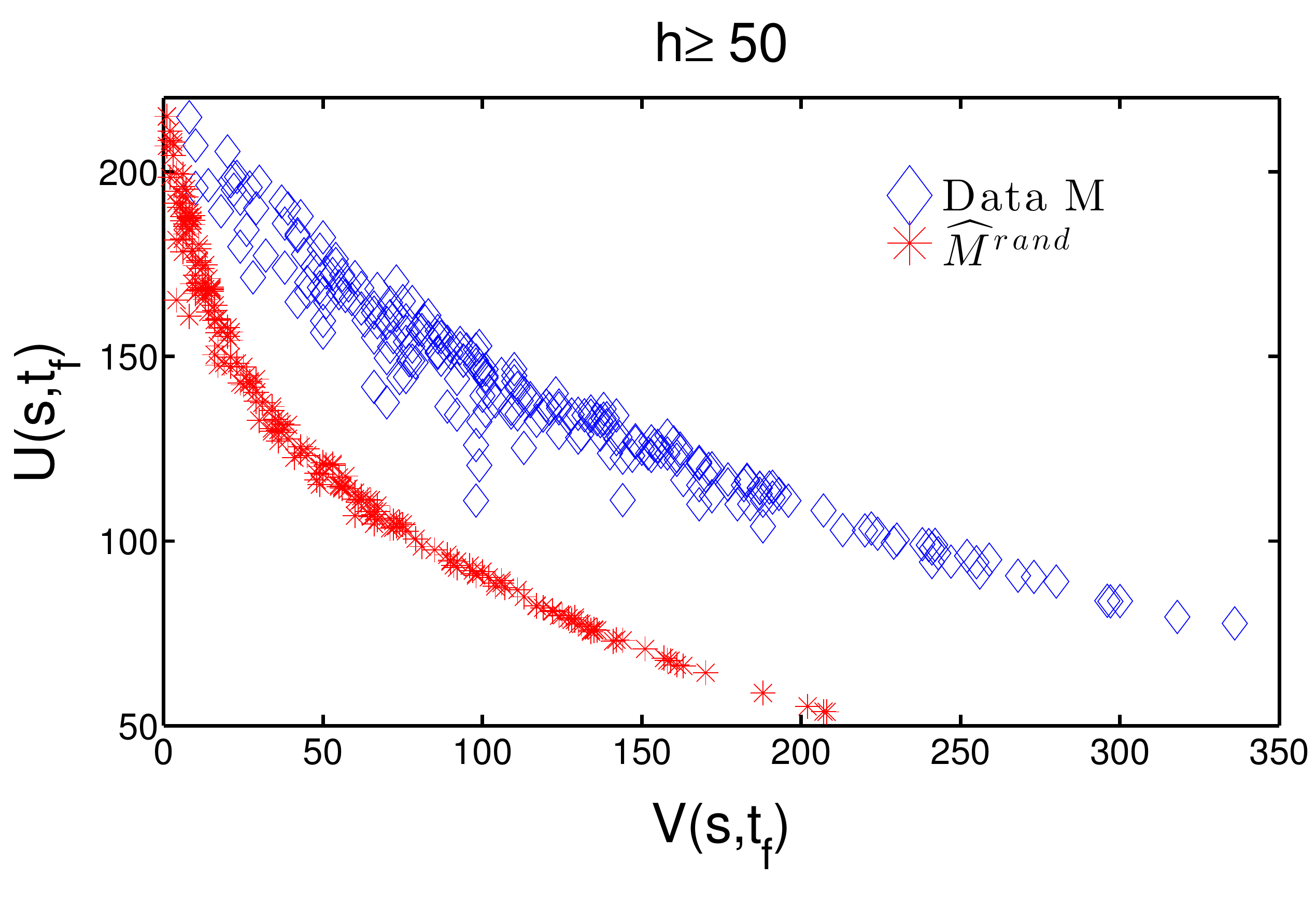}
  \includegraphics[width=0.45\textwidth]{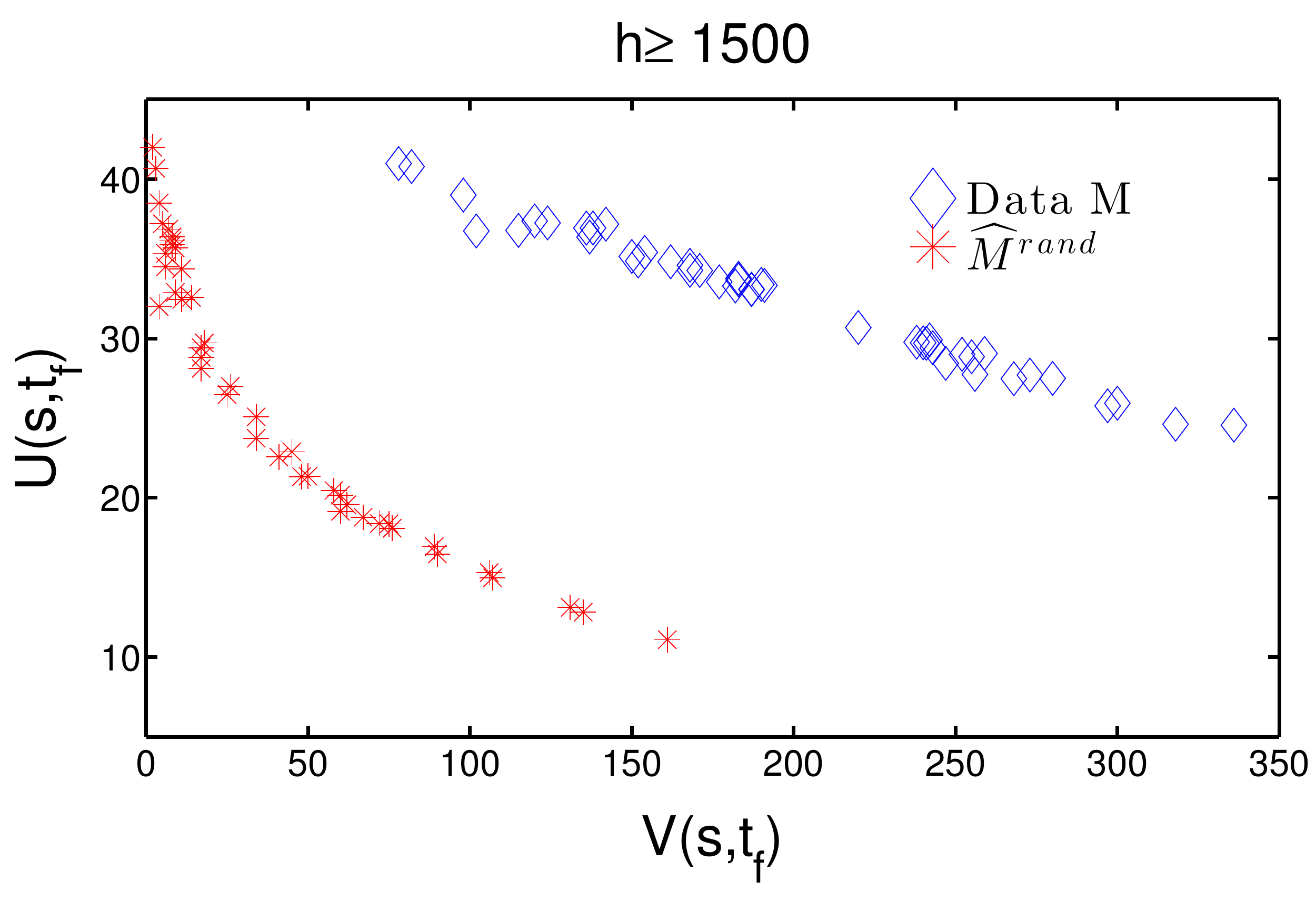}
  \includegraphics[width=0.44\textwidth]{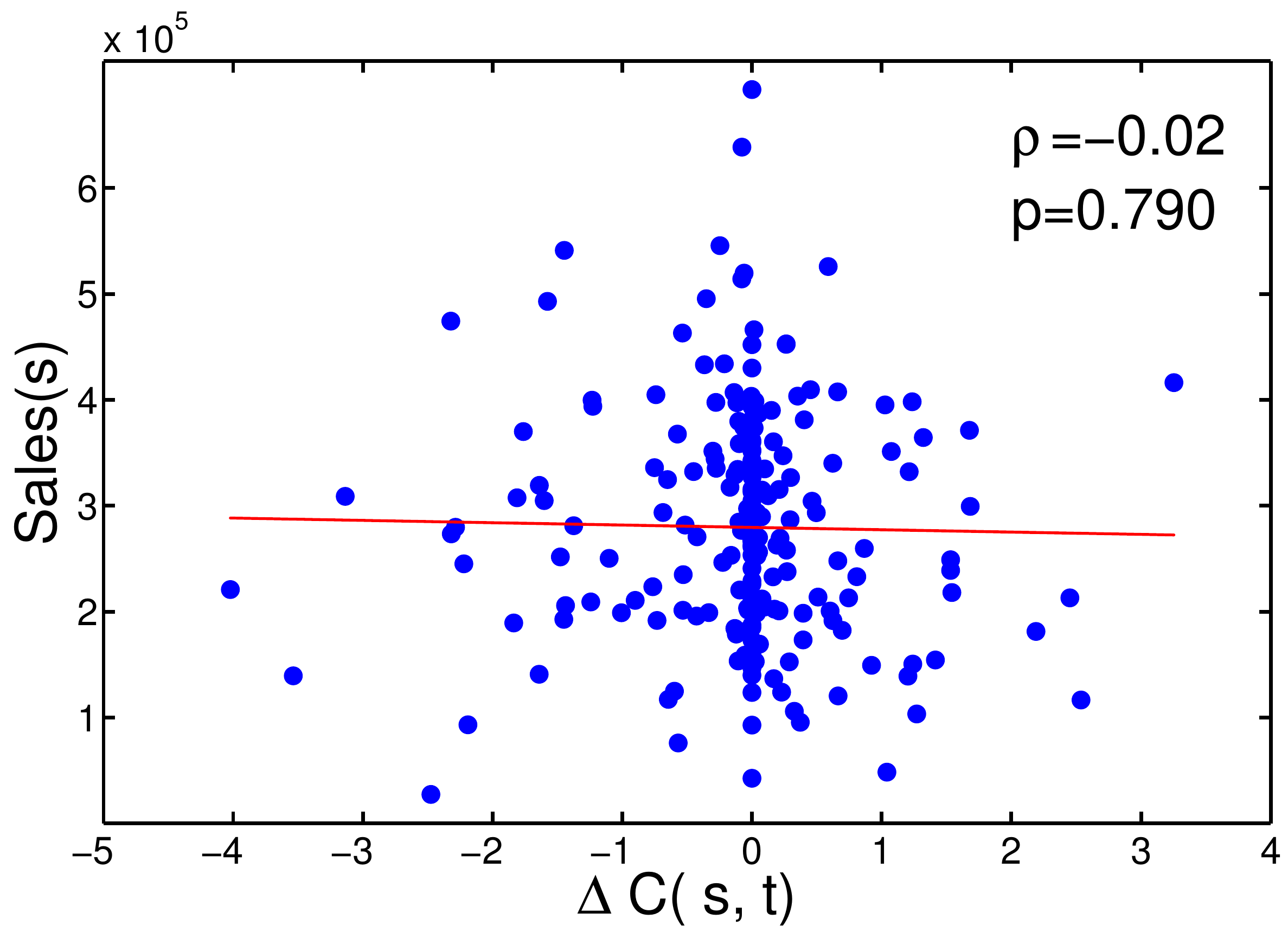}
  \includegraphics[width=0.44\textwidth]{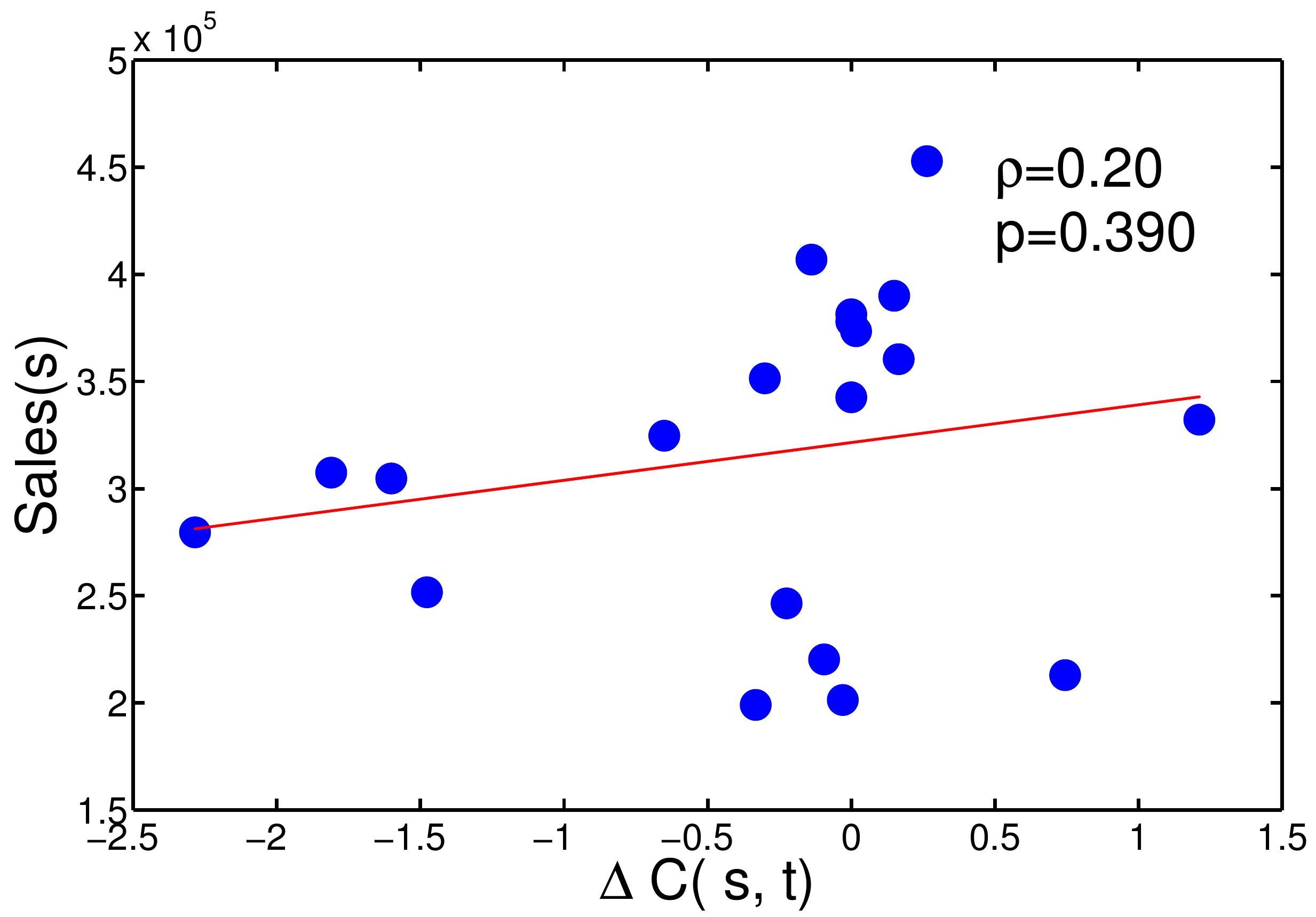}
   \caption
  {{\it Top row:} The distribution of styles in the $V$-$U$-plane computed from the randomized model $\widehat M^{rand}$ shows that the negative correlation between $V$ and $U$ is preserved for both threshold vales, $h=50$ (left) and $h=1500$ (right). However, in particular the uniformity values of styles are much smaller under randomization, when compared to the data. {\it Bottom row:} There is no correlation between complexity change and change in sales numbers for both thresholds, $h=50$ and $h=1500$.}
  \label{fig:Model3}
\end{figure}

Results for the distribution of variety and uniformity of music styles for data and for the randomized model production network $\widehat M^{rand}$ are shown in Fig. S\ref{fig:Model3} for two threshold vales, $h=50$ and $h=1500$. 
The randomization $\widehat M^{rand}$ also shows a negative correlation between $V(s,t)$ and $U(s,t)$, but especially the uniformity values are strongly underestimated in the randomized model when compared to the data.
The correlation between complexity change and sales numbers is destroyed by the randomization in $\widehat M^{rand}$, as is shown in the bottom row in Fig. S\ref{fig:Model3}.

\end{document}